\documentclass[12pt]{article}
\usepackage{graphicx}

\newcommand\pubnumber{SNSN-323-63}
\newcommand\pubdate{\today}
\usepackage{url}
\def\frascati{Laboratori Nazionali di Frascati dell'INFN\\
Via E. Fermi, 40, I-00044 Frascati, ITALY}

\def\Title#1{\begin{center} {\Large #1 } \end{center}}
\def\Author#1{\begin{center}{ \sc #1} \end{center}}
\def\Address#1{\begin{center}{ \it #1} \end{center}}

\newcommand\pubblock{\rightline{\begin{tabular}{l} \pubnumber\\
         \pubdate  \end{tabular}}}
\newenvironment{Abstract}{\begin{quotation}  }{\end{quotation}}
\newenvironment{Presented}{\begin{quotation} \begin{center} 
             PRESENTED AT\end{center}\bigskip 
      \begin{center}\begin{large}}{\end{large}\end{center} \end{quotation}}

\def\beq{\begin{equation}}
\def\eeq#1{\label{#1}\end{equation}}
\def\eeqn{\end{equation}}

\def\beqa{\begin{eqnarray}}
\def\eeqa#1{\label{#1}\end{eqnarray}}
\def\eeqan{\end{eqnarray}}

\let\bar=\overbar

\def\Dslash{\not{\hbox{\kern-4pt $D$}}}
\def\dslash{\not{\hbox{\kern-2pt $\del$}}}

\def\msb{{\bar{\ssstyle M \kern -1pt S}}}

\def\support{\footnote{For the NA62 collaboration.}}

\begin{document}
\begin{titlepage}
\pubblock

\vfill
\Title{The NA62 experiment at CERN: status and perspectives}
\vfill
\Author{ Tommaso Spadaro\support}
\Address{\frascati}
\vfill
\begin{Abstract}
The future program of the NA62 experiment at CERN SPS is currently in advanced stage of development. 
The main goal of the experiment is to measure the branching ratio of the ultra-rare decay $K^+\to\pi^+\nu\overline{\nu}$, by detecting approximately 80 events with a background
on the order of 10\%. In the present paper, the motivation behind this measurement and the overall design strategy of the experiment are briefly outlined. 
The experiment's construction status is discussed and perspectives are given for the first runs with the new detector.
\end{Abstract}
\vfill
\begin{Presented}
CKM workshop\\
Warwick, UK,  September 6--10, 2010
\end{Presented}
\vfill
\end{titlepage}
\def\thefootnote{\fnsymbol{footnote}}
\setcounter{footnote}{0}

\section{The physics case}
\begin{figure}[htb]
\centering
\includegraphics[height=3.0in]{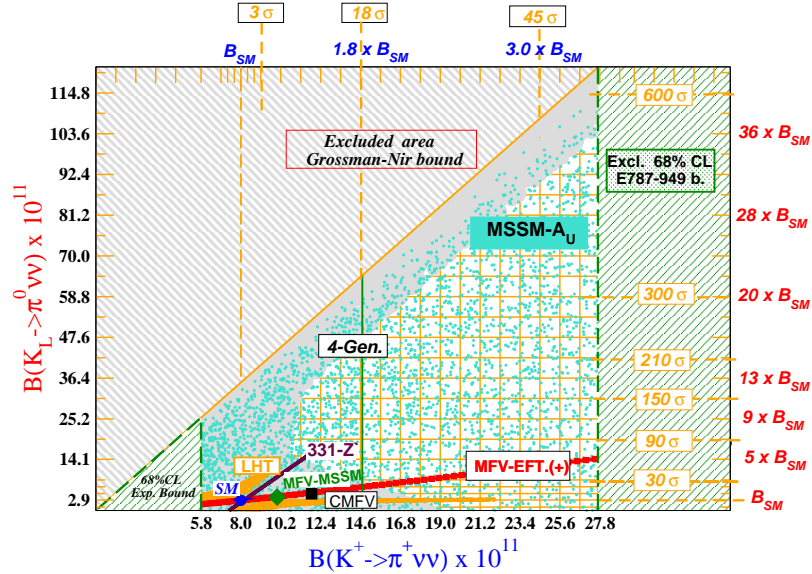}
\caption{Expected BR for $K\to\pi\nu\overline{\nu}$ modes in various new physics scenarios, from the FlaviaNet Kaon working group~\cite{mescia_flavia}.}
\label{fig:mescia}
\end{figure}

In the search for new physics (NP) effects beyond the Standard Model (SM), flavor-changing neutral-current processes are particurarly relevant. These processes are dominated by penguin and box diagrams and
can sensitively test various NP scenarios. Kaons rare and ultra-rare decays, such as $K_L\to\pi^0 l^+l^-$ and $K\to\pi\nu\overline{\nu}$, are particularly clean in that 
no long-distance contributions from processes with intermediate photons are involved and hadronic matrix elements can be obtained from branching ratios (BR) of leading $K$ decays, such as
$K\to\pi e\nu$, via isospin rotation~\cite{buraas,isidori,buchalla,MesciaSmith}. In the SM, the theoretical expectation for the charged mode~\cite{BrodGorbahn}, 
$\mathrm{BR}(K^+\to\pi^+\nu\overline{\nu})=(8.5\pm0.7)\times10^{-11}$, has a 7\% non-parametric error. In contrast, possible NP contributions could change the BR by up to a factor of three in many scenarios~\cite{mescia}. Evidence of new physics may be seen in $K\to\pi\nu\overline{\nu}$ decays even in absence of significant signals from $B$ decays 
\cite{NP_K_not_B}. Moreover, simultaneous BR measurements for $K_L\to\pi^0\nu\overline{\nu}$ and $K^+\to\pi^+\nu\overline{\nu}$ give information on CKM matrix elements in a complementary and independent manner with respect to the inputs from $B$ physics. 
At present, seven $K^+\to\pi^+\nu\overline{\nu}$ events have been identified by the BNL E949/E787 stopped-kaon decay experiments~\cite{E949_E787}. 
The measured BR is compatible with the SM prediction, although with a large uncertainty: $BR=(1.73^{+1.15}_{-1.05})\times10^{-10}$. There is still plenty of room for possible NP effects, see Fig.~\ref{fig:mescia}
from Ref.~\cite{mescia_flavia}.

\section{The NA62 experimental technique}
\begin{figure}[htb]
\centering
\includegraphics[height=2.4in]{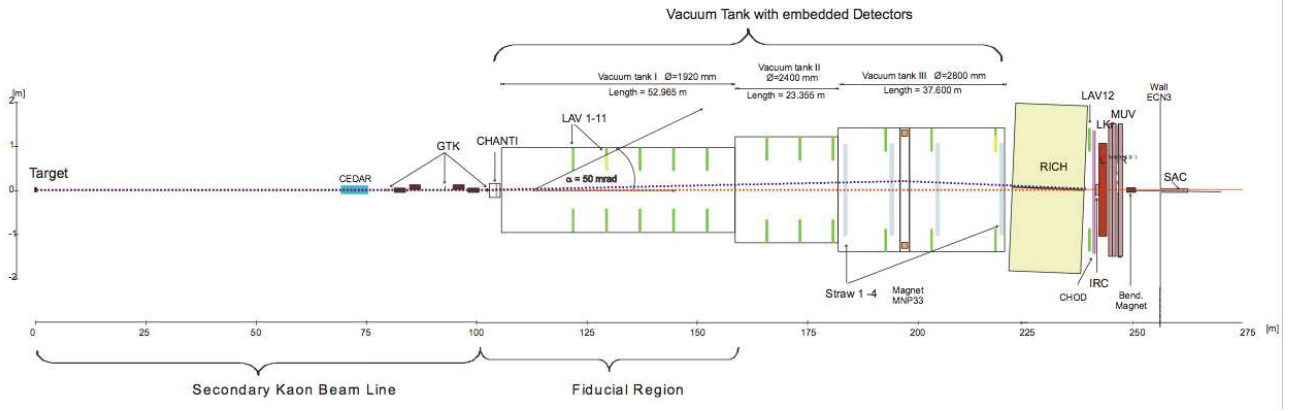}
\caption{Schematic view of the NA62 detector.}
\label{fig:NA62det}
\end{figure}

The aim of the NA62 experiment~\cite{NA62TP} is to detect about 80 $K^+\to\pi^+\nu\overline{\nu}$ events with a ${\cal O}(\sim10\%)$ signal acceptance and a background on the order of 10~\% in two years of data-taking. 
The design is inspired by years of experience 
with the NA48 apparatus and infrastructure. For NA62, the K12 beamline at the CERN SPS will be upgraded to increase the intensity by a factor of 50. In the final setup, a 400-GeV SPS primary proton beam interacts into a beryllium target and produces an unseparated 75-GeV, 800-MHz beam with 
$\sim6\%$ $K^+$, corresponding to $\sim5$~MHz kaon decays in a 60-m long fiducial volume. A transverse schematic view of the NA62 detector is shown in Fig.~\ref{fig:NA62det}.

The guiding principles in the experiment design follow from the need to sustain a high-rate environment while guaranteeing
high-resolution timing. The goal is to identify a signal BR of $\sim10^{-10}$ with a total background rejection of the order of $10^{12}$ against the leading $K^+$ decay modes.

Two- and three-body decay modes will be reduced by a factor of $\sim10^4$ by cutting on the missing mass of reconstructed candidates.
For this purpose, a fast up-stream tracker of every particle in the beam, the so-called Gigatracker (see Sec.~\ref{sec:GT}), is used to measure incoming K momentum. Downstream to a 60-m long fiducial region for $K$ decays, 
a straw-chamber magnetic spectrometer is used to measure with high resolution daughter particle momenta.

Further rejection of $K_{\mu2,3,4}$ and $K_{e2,3,4}$ background will be obtained with a ring-imaging Cerenkov counter (see Sec.~\ref{sec:RICH}), used to efficiently and non-destructively identify daughter pions
from muons and electrons. The $\pi$/$\mu$ separation is critical to achieve sufficient rejection for $K_{\mu2}$ decays. For this purpose, additional information will be provided by a sampling calorimeter, the so-called muon veto, placed after the 27~$X_0$'s of the existing LKr NA48
electromagnetic calorimeter. 

Rejection of background from nuclear interactions of charged beam particles other than $K^+$ will be guaranteed by a differential Cerenkov counter, the so-called CEDAR, placed before
$K^+$'s enter the decay region.

Rejection of modes with $\pi^0$'s and/or (possibly radiative) photons will be provided by a hermetic, high-efficiency photon-veto system, covering from 0-50~mrad $\gamma$ 
emission angles. This has to provide a rejection factor of $10^8$ against $K^+\to\pi^+\pi^0$. 

\subsection{Fast Tracking}
\label{sec:GT}
The Gigatracker~\cite{IEEE_GT} will measure time of individual beam particles. It will also provide a momentum measurement for single particles with 
precision much better than that given by the momentum bite of the beam. This system is placed upstream, just before the decay volume, and must sustain rates of up to
800~MHz. Three silicon micro-pixel stations with a total thickness of less than 0.5\%~$X_0$ each will provide position measurement while particles traverse a magnetic achromat.
A total of 18\,000, $300\times300~\mu\mathrm{m}^2$ pixels for a sensitive area of $60\times27$~mm$^2$ will provide spatial hit resolution
of $\sim100~\mu$m. Momentum will be measured with a fractional error of $\sim0.2\%$, corresponding to 150~MeV resolution, while direction will be determined with 12~$\mu$rad angular resolution. In order not to cause station-by-station hit mismatch in more than 1\% of the cases, a hit time resolution better than 200~ps is required. The read out has to 
sustain rates of up to 150~KHz. The R\&D is almost completed, with
two read out prototypes developed and compared, both with FE circuits in 130-nm IBM CMOS technology.

\subsection{PID of decay products}
\label{sec:RICH}
\begin{figure}[htb]
\centering
\includegraphics[height=2.in]{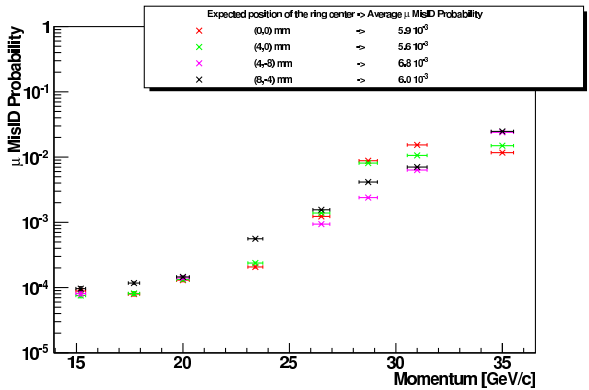}
\caption{Result from a RICH prototype beam test: muon mis-identification probability as a pion as a function of particle momentum. 
To estimate the effect induced by the use of a fixed ring position in the prototype, 
measurements were repeated comparing four alignment positions of the mirror, corresponding to different colors in the plot.}
\label{fig:RICH}
\end{figure}

A ring-imaging Cerenkov counter, the RICH~\cite{RICH}, will provide rejection for muons with less than 0.5\% mis-ID probability for events not identified by the muon veto. More than three
standard deviations of $\pi/\mu$ separation should be achieved in the $K\to\pi\nu\overline{\nu}$ pion momentum range, $15<p_\pi<35$~GeV.
Time determination with a resolution better than 100~ps should be guaranteed, to efficiently match with Gigatracker information. 
This performance will be obtained bu using a 17-m long, 3-m diameter volume,
filled with 1~atm Ne gas acting as Cerenkov radiator. 
Mirrors at the downstream side of the volume will focus rings of Cerenkov light into two separated regions on the upstream side. 
These are instrumented with 2000 photomultiplier tubes (PMT's), each 18-mm wide. In a dedicated test beam for a prototype with $\sim400$ PMT's a muon rejection better than
1\% has been measured, with an overall pion loss of few per mil (see fig~\ref{fig:RICH}) 
and a time resolution better than 100~ps, these figures holding across the momentum range of interest.

\subsection{Efficient photon vetoing}
\label{sec:PHVETO}
A system of different detectors will veto photons and will provide a rejection of $10^{8}$ for photons from $K^+$ decay in a 60~m-long fiducial volume, allowing the background from  $K^+\to\pi^+\pi^0$ decays to be
reduced to less than one part in $10^{12}$. Photons emitted at very small angle, $<\sim2$~mrad, will be detected by compact calorimeters
in the forward direction, with a required inefficiency of $<10^{-6}$ above 6~GeV. In the angular range between 1~mrad and 8~mrad, the existing NA48 LKr calorimeter will be re-used,
profiting of a measured inefficiency $<10^{-5}$ for photons above 6~GeV. At large angle, between 8~mrad and 50~mrad, a new system (so-called LAV) will provide $\gamma$ detection
with an inefficiency $<\sim10^{-4}$ above 100~MeV.
\subsubsection{The LAV system}
\begin{figure}[htb]
\centering
\includegraphics[height=2.in]{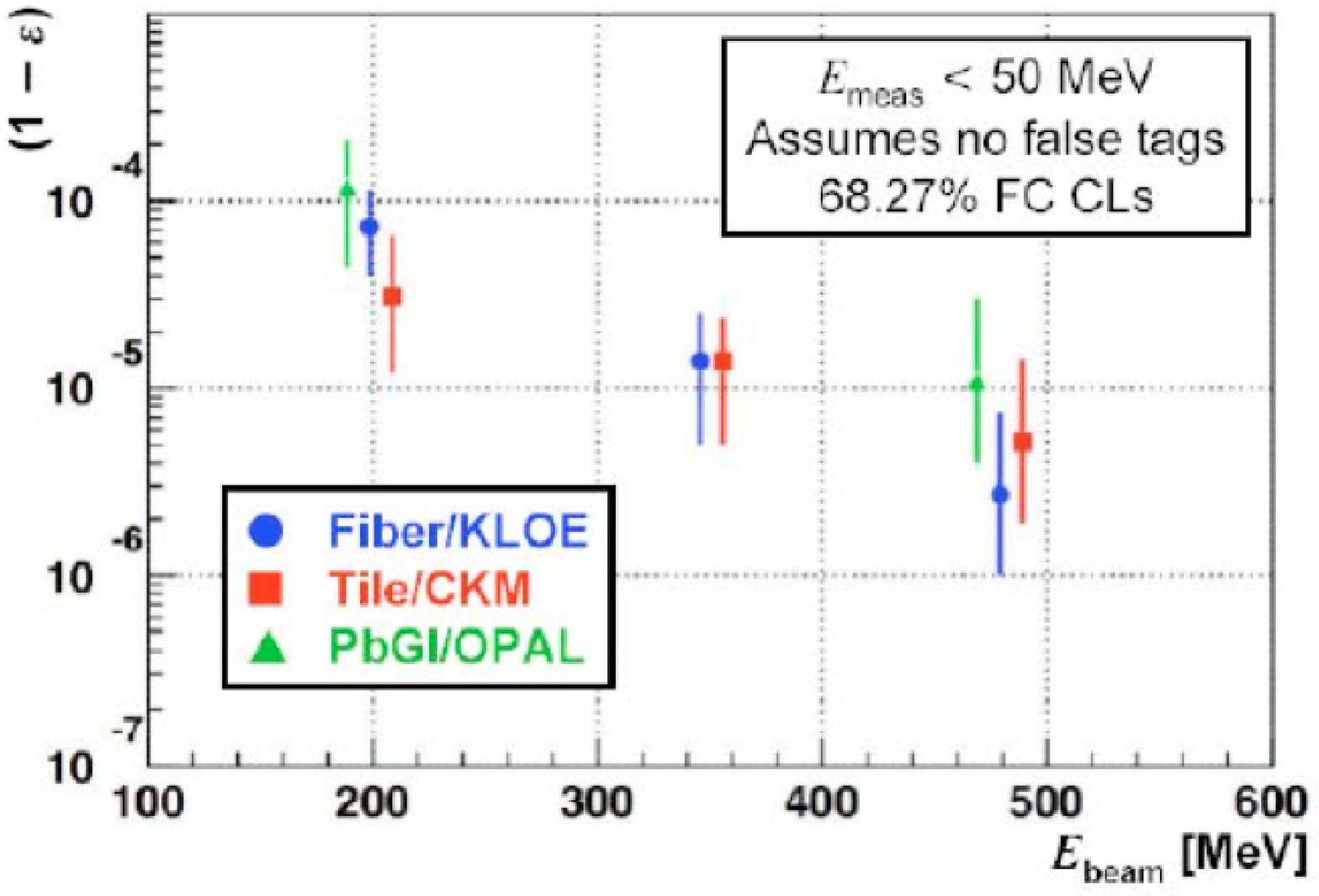}
\includegraphics[height=2.2in]{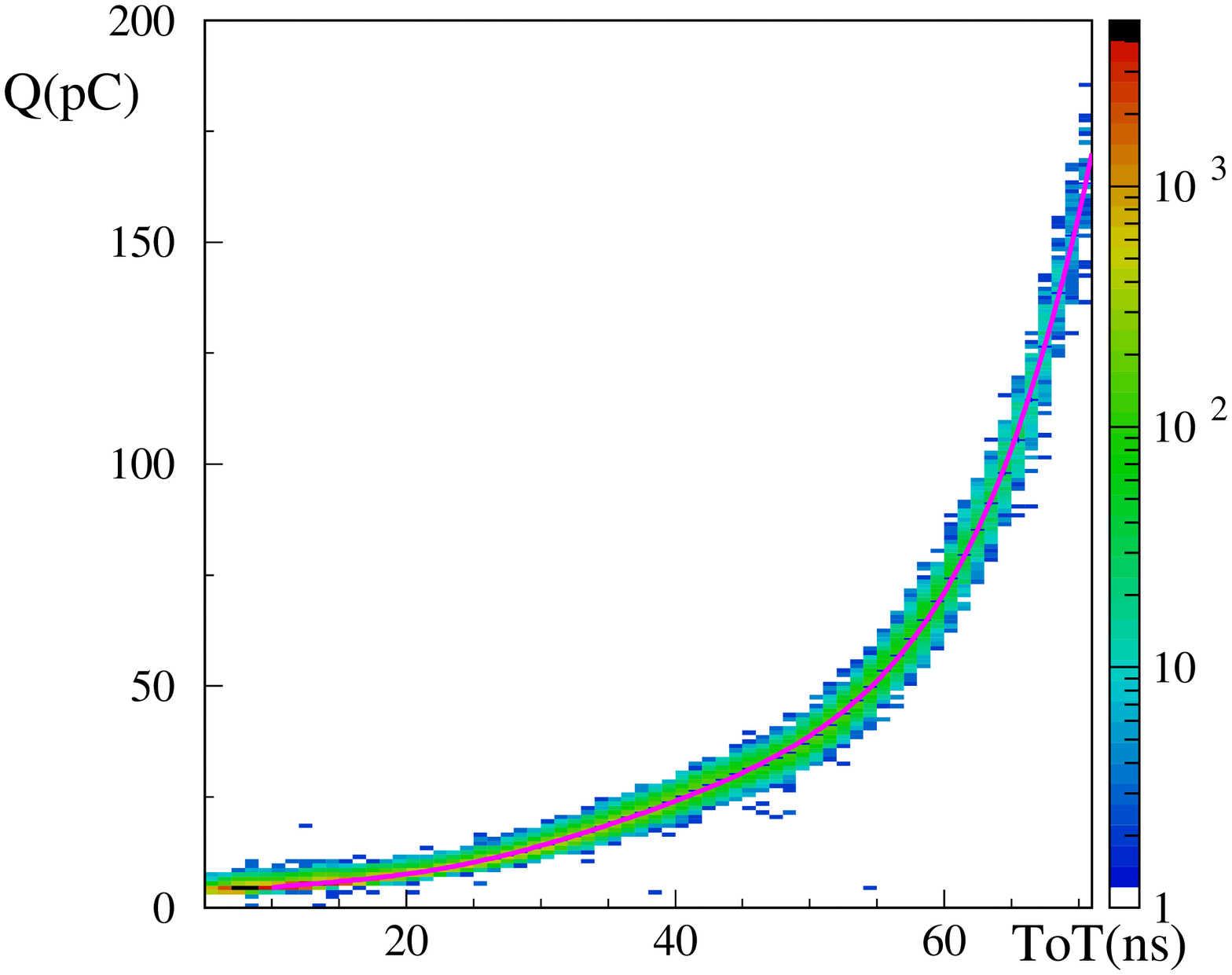}
\caption{Left: results from large-angle photon veto prototype beam tests. Inefficiency for electron detection as a function of particle momentum is compared for a fiber calorimeter
``\`a-la KLOE'' (circles), for a structure alternating scintillator and lead tiles, ``\`a-la CKM'' (squares), and for a structure made of OPAL lead-glass blocks (triangles).
Right: data from test with electron beam. Charge integrated by a QDC with 0.1~pC LSB versus time over a 30~mV threshold, meausred with a prototype of the LAV front-end electronics and
a TDC with 0.25~ps LSB.}
\label{fig:VETOtb}
\end{figure}

After an intense R\&D activity, the re-use of SF57 lead glass blocks from the dismounted OPAL barrel electromagnetic calorimeter, already instrumented with R-2238 Hamamatsu
phototubes, has been validated. The inefficiency measured with dedicated test beams 
satisfies the requirements and is comparable with other alternatives, including lead/scintillating-fiber or lead/scintillating-tile sampling calorimeters (see Fig.~\ref{fig:VETOtb}, left).

The LAV will be made of 12 stations of increasing diameter to cover hermetically the angular range from 7 to 50~mrad. Each station will be composed of four
or five layers, for a total depth of 29 to 37 $X_0$'s. Layers are staggered to guarantee that incident particles must encounter at least three blocks, corresponding to more than 20 $X_0$'s. With
32 to 48 crystals per layer, a total of $\sim2500$ blocks will be used. Since high sensitivity to photons in the range from 20~MeV to 20~GeV is required,
the front end electronics must guarantee a wide dynamic range. Since the 
typical yield is $\sim0.3$~photo-electrons per MeV and the typical PMT gain is $10^6$,
signals from 20~mV up to 10~V on a 50~$\Omega$ load must be treated. A simple and cost-effective solution, easy to scale and to integrate with a common NA62 trigger and data acquisition infrastructure has been adopted:
a time-over-threshold discriminator, with multiple adjustable thresholds. Signals will be clamped, split into two, amplificated, and discriminated with two thresholds to allow slewing
corrections. The digital output in LVDS standard will allow accurate leading and trailing edge time determinations. From the time-over-threshold, a $10\%$ resolution measurement of
the charge will be made (see Fig.~\ref{fig:VETOtb}, right), allowing the LAV system to operate as a calorimeter as well as a veto.
Test beam results show that a fractional energy resolution $\sim10\% E(\mathrm{GeV})^{-1/2}$ and a time resolution $\sim300~\mathrm{ps}E(\mathrm{GeV})^{-1/2}$ are achieved.
 
\subsection{Experiment status}
In September 2005, the experiment was presented to the CERN SPS Committee and in December 2005 the R\&D was endorsed by the CERN Research Board. In December 2008, 
the experiment was approved by the 
CERN Research Board. At present, the collaboration has 191 participants from 25 institutes. Beam tests have been performed for advanced prototypes or parts 
of single sub-detectors, as discussed above. 
Construction and commissioning will last until 2010, and the first physics run is expected to take place in 2013.

The experiment aims to collect $\sim80$ signal events in two years of data taking, with a ${\cal O}(\sim10\%)$ signal acceptance and a total background of less than 10\%.

\end{document}